\documentclass[conference]{IEEEtran}
\IEEEoverridecommandlockouts

\usepackage{hyperref}

\usepackage{tikz}
\usepackage{amsmath}
\usepackage{todonotes}
\usepackage{booktabs}
\usepackage[inline]{enumitem}
\usepackage{threeparttable}
\usepackage{etoolbox}
\usepackage{flushend}
\appto\TPTnoteSettings{\footnotesize}

\newcommand{\ctab}{\hspace{1em}}

\newcommand{\JS}{JavaScript}
\newcommand{\EL}{EasyList}
\newcommand{\EP}{EasyPrivacy}
\newcommand{\ABP}{AdBlock Plus}
\newcommand{\UBO}{uBlock Origin}
\newcommand{\PG}{PageGraph}
\newcommand{\SG}{subgraph}
\newcommand{\etal}{\textit{et al.}}
\newcommand{\IE}{i.e.,}
\newcommand{\EG}{e.g.,}
\newcommand{\CrawlStartDate}{Oct 23, 2019}
\newcommand{\CrawlEndDate}{Oct 24, 2019}
\newcommand{\CrawlRespondingDomains}{88,035}
\newcommand{\CrawlNotRespondingDomains}{11,965}
\newcommand{\CrawlPageGraphDomains}{87,941}

\newcommand{\CrawlNonPageGraphDomains}{4,286}
\newcommand{\CrawlFilterListDate}{Nov 2, 2019}

\newcommand{\SignatureMinEdges}{13}
\newcommand{\SignatureMinNodes}{4}

\newcommand{\NumSignatures}{1,995,444}
\newcommand{\NumSignaturesUnique}{400,166}

\newcommand{\NumBlockedSignatures}{2,001}
\newcommand{\NumSmallSignatures}{3,472}
\newcommand{\NumBlockedAndSmallSignatures}{5,473}

\newcommand{\NumFingerprintedScriptsUnique}{14,801}            
\newcommand{\NumBlockedFingerprintedScriptsUnique}{11,212}     
\newcommand{\NumBlockedFingerprintedScriptsInstances}{68,278}  
\newcommand{\NumNewFingerprintedScriptsUnique}{3,589}          
\newcommand{\NumNewFingerprintedScriptsInstances}{12,044}      
\newcommand{\NumNewFingerprintedScriptURLs}{3,091}             
\newcommand{\NumNewFingerprintedScriptURLInstances}{11,546}    
\newcommand{\PctMoreNewFingerprintedScriptURLs}{27.57\%}       

\newcommand{\NumSmallFingerprintedScriptsUnique}{195,727}            
\newcommand{\NumSmallBlockedFingerprintedScriptsUnique}{45,327}      
\newcommand{\NumSmallBlockedFingerprintedScriptsInstances}{145,500}  
\newcommand{\NumSmallNewFingerprintedScriptsUnique}{150,400}         
\newcommand{\NumSmallNewFingerprintedScriptURLs}{82,483}             
\newcommand{\NumSmallNewFingerprintedScriptURLInstances}{133,153}    

\newcommand{\NumDomainsHostingFingerprintedScripts}{2,873}

\newcommand{\NumDomainsHostingNewFingerprintedScripts}{1,965}
\newcommand{\NumDomainsMatchGroundTruthSigs}{56,390}

\newcommand{\NumWebsitesWithEvasion}{10,973}
\newcommand{\PctWebsitesWithEvasion}{12.48\%}

\newcommand{\NumGAEvasionsUnique}{125}
\newcommand{\PctGAEvasionsUnique}{17.36\%}
\newcommand{\NumGAEvasionInstances}{5,283}
\newcommand{\PctGAEvasionInstances}{66.67\%}

\newcommand{\NumInstanceEvasionByMovedScript}{7,924}
\newcommand{\NumUniqeScriptsEvasionByMovedScript}{720}
\newcommand{\PctInstanceEvasionByMovedScript}{65.79\%}
\newcommand{\PctUniqueEvasionByMovedScript}{20.06\%}
\newcommand{\PctInstanceEvasionNotByMovedScript}{34.21\%}

\newcommand{\NumInstanceEvasionByBundling}{117}
\newcommand{\NumEvasionByBundling}{85}
\newcommand{\PctInstanceEvasionByBundling}{0.97\%}
\newcommand{\PctUniqueEvasionByBundling}{13.88\%}

\newcommand{\NumSmallEvasionByInlining}{67,917}
\newcommand{\NumEvasionByInlining}{498}
\newcommand{\PctInstanceEvasionByInlining}{4.13\%}
\newcommand{\PctUniqueEvasionByInlining}{2.37\%}
\newcommand{\NumWebsitesEvasionByInlining}{231}

\newcommand{\NumInstanceEvasionByCommonCode}{3,505}
\newcommand{\NumEvasionByCommonCode}{2,286}
\newcommand{\PctInstanceEvasionByCommonCode}{29.10\%}
\newcommand{\PctUniqueEvasionByCommonCode}{63.69\%}

\newcommand{\NumFilterListRules}{586}  

\newcommand{\VEightScriptIdURL}{\url{https://v8docs.nodesource.com/node-0.8/d0/d35/classv8_1_1_script.html}}
\newcommand{\JSMicrotasksURL}{\url{https://javascript.info/microtask-queue}}
\newcommand{\PageGraphWikiURL}{\url{https://github.com/brave/brave-browser/wiki/PageGraph}}

\begin{document}

\date{}

\title{Improving Web Content Blocking With Event-Loop-Turn Granularity JavaScript Signatures}

\author{}

\author{
\IEEEauthorblockN{Quan Chen}
\IEEEauthorblockA{North Carolina State University\\
qchen10@ncsu.edu}
\and
\IEEEauthorblockN{Peter Snyder}
\IEEEauthorblockA{Brave Software\\
pes@brave.com}
\and
\IEEEauthorblockN{Ben Livshits}
\IEEEauthorblockA{Brave Software\\
ben@brave.com}
\and
\IEEEauthorblockN{Alexandros Kapravelos}
\IEEEauthorblockA{North Carolina State University\\
akaprav@ncsu.edu}
} 

\maketitle

\begin{abstract}
Content blocking is an important part of a performant, user-serving,
privacy respecting web.  Current content blockers work by building trust
labels over URLs.  While useful, this approach has many well understood
shortcomings.  Attackers may avoid detection by changing URLs or domains,
bundling unwanted code with benign code, or inlining code in pages.

The common flaw in existing approaches is that they evaluate code based on its
delivery mechanism, not its behavior. In this work we address this problem by
building a system for generating signatures of the privacy-and-security
relevant behavior of executed \JS{}.  Our system uses as the unit of analysis each
script's behavior during each turn on the \JS{} event loop.  Focusing on
event loop turns allows us to build highly identifying signatures for \JS{}
code that are robust against code obfuscation, code bundling, URL
modification, and other common evasions, as well as handle unique aspects
of web applications.

This work makes the following contributions to the problem of measuring and
improving content blocking on the web: First, we design and implement a novel
system to build per-event-loop-turn signatures of \JS{} behavior through deep
instrumentation of the Blink and V8 runtimes.  Second, we apply these
signatures to measure how much privacy-and-security harming code
is missed by current content blockers, by using \EL{} and \EP{} as
ground truth and finding scripts that have the same privacy and security
harming patterns.  We build \NumSignatures{} signatures of
privacy-and-security relevant behaviors from
\NumBlockedFingerprintedScriptsUnique{} unique scripts blocked by filter lists,
and find \NumNewFingerprintedScriptsUnique{} unique scripts hosting known
harmful code, but missed by filter lists, affecting \PctWebsitesWithEvasion{}
of websites measured. Third, we provide a taxonomy of ways scripts avoid
detection and quantify the occurrence of each.  Finally, we present defenses
against these evasions, in the form of filter list additions where possible,
and through a proposed, signature based system in other cases.

As part of this work, we share the implementation of our signature-generation
system, the data gathered by applying that system to the Alexa 100K, and
\NumFilterListRules{} \ABP{} compatible filter list rules to block instances
of currently blocked code being moved to new URLs.
\end{abstract}

\section{Introduction}
\label{sec:intro}

Previous research has documented the many ways content blocking tools
improve privacy, security, performance, and user experience online
(\EG{} \cite{miroglio2018effect, garimella2017ad, li2012knowing, pujol2015annoyed}).
These tools are the current stage in a long arms race between communities that
maintain privacy tools, and online trackers who wish to evade them.

Initially, communities identified domains associated with tracking, and
generated hosts files that would block communication with these
undesirable domains. Trackers, advertisers, and attackers reacted by moving
tracking resources to domains that served both malicious and user-serving code,
circumventing host based blocking. In response, content blocking communities
started identifying URLs associated with undesirable code, to distinguish
security-and-privacy harming resources from user-desirable ones, when both were
\textit{served from the same domain}, such as with content delivery networks
(CDNs).

URL-based blocking is primarily achieved through the crowd-sourced generation of
filter lists containing regular-expression style patterns that determine which
URLs are desirable and which should be blocked (or otherwise granted less
functionality). Popular filter lists include \EL{} (EL) and \EP{} (EP). The
non-filter-list based web privacy and security tools (\EG{} Privacy Badger,
NoScript, etc.) also use URL or domain-level determinations when making access
control decisions.

However, just as with hosts-based blocking, URL-based blocking has several
well known weaknesses and can be easily circumvented.
Undesirable code can be moved to one-off, rare URLs, making crowdsourced
identification difficult. Furthermore, such code can be mixed with benign code
in a single file, presenting content blockers with a lose-lose choice between
allowing privacy or security harm, or a broken site.  Finally, unwanted code can
also be ``inlined'' in the site (\IE{} injected as text into a \texttt{<script>}
element), making URL level determinations impossible.

Despite these well known and simple circumventions, the privacy and research
community lacks even an understanding of the scale of the problem, let alone
useful, practical defenses.  Put differently, researchers and activists know
they \textit{might} be losing the battle against trackers and online attackers,
but lack measurements to determine if this is true, and if so, by how much.
Furthermore, the privacy community lacks a way of providing practical (\IE{}
web-compatible) privacy improvements that are robust regardless of how the
attackers choose to deliver their code.

Fundamentally, the common weakness in URL-based blocking tools is,
at its root, a mismatch between the targeted behavior
(\IE{} the privacy-and-security harming behavior of scripts), and
the criteria by which the blocking decisions are made (\IE{} the delivery
mechanism). This mismatch allows for straightforward evasions that are easy for
trackers to implement, but difficult to measure and defend against.

Addressing this mismatch requires a solution that is able to identify the
behaviors already found to be harmful, and base the measurement tool and/or the
blocking decisions on those behaviors. A robust solution must target a
granularity \textit{above} individual feature accesses (since decisions made at
this level lack the context to distinguish between benign and malicious feature
use) but \textit{below} the URL level (since decisions at this level lack the
granularity to distinguish between malicious and benign code delivered from the
same source). An effective strategy must target harmful behavior independent of
how it was delivered to the page, regardless of what other behavior was
bundled in the same code unit.

In this work, we address the above challenges through the design and
implementation of a system for building signatures of privacy-and-security
harming functionality implemented in \JS{}. Our system extracts script behaviors
that occur in one \JS{} event loop turn~\cite{eventloop}, and builds signatures
of these behaviors from scripts known to be abusing user privacy. We
base our ground truth of known-bad behaviors on scripts blocked by the popular
crowdsourced filter lists (\IE{} EL and EP), and generate signatures to identify
patterns in how currently-blocked scripts interact with the DOM, \JS{} APIs
(\EG{} the Date API, cookies, storage APIs, etc), and initiate network requests.
We then use these signatures of known-bad behaviors to identify the same code
being delivered from other URLs, bundled with other code, or inlined in a site.

We generate per-event-loop-turn signatures of known-bad scripts by crawling the
Alexa 100K with a novel instrumented version of the Chromium
browser. The instrumentation covers Chromium's Blink layout engine and its V8
\JS{} engine, and records script interactions with the web pages into a
graph representation, from which our signatures are then generated. We use these
signatures to both measure how often attackers evade filter lists, and as the
basis for future defenses.

In total we build \NumSignatures{} high-confidence signatures of
privacy-and-security harming behaviors (defined in Section \ref{sec:meth}) from
\NumBlockedFingerprintedScriptsUnique{} scripts blocked by \EL{} and \EP{}. We
then use our browser instrumentation and collected signatures to identify
\NumNewFingerprintedScriptsUnique{} new scripts containing
identically-performing privacy-and-security harming behavior, served from
\NumDomainsHostingNewFingerprintedScripts{} domains and affecting
\PctWebsitesWithEvasion{} of websites. Further, we use these signatures, along
with code analysis techniques from existing research, to categorize the
\textit{method} trackers use to evade filter lists.  Finally, we use our
instrumentation and signatures to generate new filter list rules for
\NumUniqeScriptsEvasionByMovedScript{} URLs that are moved instances of known
tracking code, which contribute to \PctInstanceEvasionByMovedScript{} of all
instances of filter list evasion identified by our approach, and describe how
our tooling and findings could be used to build defenses against the rest of the
\PctInstanceEvasionNotByMovedScript{} instances of filter list evasions.

\subsection{Contributions}
This work makes the following contributions to improving the state of web
content blocking:

\begin{enumerate}
  \item The \textbf{design and implementation} of a system for generating
  signatures of \JS{} behavior.  These signatures are
  robust to popular obfuscation and \JS{} bundling tools and rely on
  extensive instrumentation of the Blink and V8 systems.

  \item A \textbf{web-scale measurement of filter list evasion}, generated
  by measuring how often privacy-sensitive behaviors of scripts labeled
  by \EL{} and \EP{} are repeated by other scripts in the Alexa 100K.

  \item A \textbf{quantified taxonomy of filter list evasion techniques}
  generated by how often scripts evade filter lists by changing URLs,
  inlining, or script bundling.

  \item \NumFilterListRules{} new \textbf{filter list rules} for identifying
  scripts that are known to be privacy-or-security related, but evade existing
  filter lists by changing URLs.
\end{enumerate}

\subsection{Research Artifacts and Data}
As part of this work we also share as much of our research outcomes and
implementation as possible.  We share the source of our Blink and V8
instrumentation, along with build instructions.  Further, we share our complete
dataset of applying our \JS{} behavior signature generation pipeline to the
Alexa 100K, including which scripts are encountered, the execution graphs
extracted from each measured page, and our measurements of which scripts are (or
include) evasions of other scripts.

Finally, we share a set of \ABP{} compatible filter list additions to block
cases of websites moving existing scripts to new URLs (\IE{} the subset of the
larger problem that \textit{can} be defended against by existing
tools)~\cite{semanticsignatures}. We note many of these filter list additions
have already been accepted by existing filter list maintainers, and note those
cases.

\section{Problem Area}
\label{sec:problem}

This section describes the evasion techniques that existing
content blocking tools are unable to defend against, and which the rest
of this work aims to measure and address.

\subsection{Current Content Blocking Focuses on URLs}
\label{sec:problem:current-tools}
Current content-blocking tools, both in research and popularly
deployed, make access decisions based on URLs. \ABP{} and \UBO{}, for example,
use crowd-sourced filter lists (i.e. lists of regex-like patterns) to
distinguish trusted from untrusted URLs.

Other content blockers make decisions based on the domain of a resource, which
can be generalized as broad rules over URLs. Examples of such tools include
Ghostery, Firefox's ``Tracking Protection'', Safari's ``Intelligent Tracking
Protection'' system, and Privacy Badger.  Each of these tools build trust labels
over domains, though they differ in both how they determine those labels
(expert-curated lists in the first two cases, machine-learning-like heuristics
in the latter two cases), and the policies they enforce using those domain
labels.

Finally, tools like NoScript block all script by default, which conceptually
is just an extremely general, global trust label over all scripts. NoScript
too allows users to create per-URL exception rules.

\subsection{URL-Targeting Systems Are Circumventable}
Relying on URL-level trust determinations leaves users vulnerable to practical,
trivial circumventions. These circumventions are common and well understood by
the web privacy and security communities.  However, these communities
lack both a way to measure the scale of the problem and deploy practical counter
measures. The rest of this subsection describes the techniques used to evade
current content-blocking tools:

\subsubsection{Changing the URL of Unwanted Code}
The simplest evasion technique is to change the URL of the unwanted
code, from one identified by URL-based blocking tools to one not identified by
blocking tools. For example, a site wishing to deploy a popular tracking script
(\EG{} \textit{https://tracker.com/analytics.js}), but which is blocked by filter
lists, can copy the code to a new URL, and reference the code there (\EG{}
\textit{https://example.com/abc.js}). This will be successful until the new URL
is detected, after which the site can move the code again at little to no cost.
Tools that generate constantly-changing URLs, or which move tracking scripts
from a third-party to the site's domain (first party) are a variation of this
evasion technique.

\subsubsection{Inlining Tracking Code}
A second evasion technique is to inline the blocked code, by inserting the code
into the text of a \texttt{<script>} tag (as opposed to having the tag's
\texttt{src} attribute point to a URL, \IE{} an external script). This process
can be manual or automated on the server-side, to keep inlined code up to date.
This technique is especially difficult for current tools to defend against,
since they lack a URL to key off.\footnote{One exception is uBlock Origin,
which, when installed in Firefox, uses non-standard API's\cite{ffFilterResponse}
to allow some filtering of inline script contents.  However, because this
technique is rare, and also trivially circumvented, we do not consider it
further in this work.}

\subsubsection{Bundling Tracking Code with Benign Code}
Trackers also evade detection by bundling tracking-related code with benign
code into a single file (\IE{} URL), and forcing the privacy tool to make
a single decision over both sets of functionality.  For example, a
site which includes tracking code in their page could combine it with other,
user-desirable code units on their page (\EG{} scripts for performing form
validation, creating animations, etc.) and bundle it all together into a single
\JS{} unit (\EG{} \textit{combined.min.js}). URL-focused tools face the
lose-lose decision of restricting the resource (and breaking the website,
from the point of view of the user) or allowing the resource (and allowing
the harm).

Site authors may even evade filter lists unintentionally. Modern websites use
build tools like WebPack\footnote{\url{https://webpack.js.org/}},
Browserify\footnote{\url{http://browserify.org/}}, or
Parcel\footnote{\url{https://parceljs.org/}} that combine many \JS{} units into
a single, optimized script.  (Possibly) without meaning to, these tools bypass
URL-based blocking tools by merging many scripts, of possibly varying
desirability, into a single file. Further, these build tools generally
``minify'' \JS{} code, or minimize the size and number of identifiers in the
code, which can further confuse naive code identification techniques.

\subsection{Problem - Detection Mismatch}
The root cause for why URL-based tools are trivial to evade is the mismatch
between what content blockers want to block (\IE{} the undesirable script
behaviours) and how content blockers make access decisions (\IE{} how the code
was delivered to the page).  Attackers take advantage of this mismatch to evade
detection; URLs are cheap to change, script behavior is more difficult to
change, and could require changes to business logic. Put differently, an
effective privacy-preserving tool should yield the same state in the browser
after executing the same code, independent of how the code was delivered,
packaged, or otherwise inserted into the document.

We propose an approach that aligns the content blocking decisions with the
behaviors which are to be blocked. The rest of this paper presents such a
system, one that makes blocking decisions based on patterns of \JS{} behavior,
and not delivery URLs.  Doing so provides both a way to measuring how often
evasions currently occur, and the basis of a system for providing better, more
robust privacy protections.

\section{Methodology}
\label{sec:meth}

This section presents the design of a system for building signatures of the
privacy-and-security relevant behavior of \JS{} code, per event loop
turn~\cite{eventloop}, when executed in a web page.  The web has a
single-threaded execution model, and our system considers the sum of behaviors
each script engages in during each event loop turn, from the time the script
begins executing, until the time the script yields control.

In the rest of this section, we start by describing why building these \JS{}
signatures is difficult, and then show how our system overcomes these
difficulties to build high-fidelity, per event-loop-turn signatures of \JS{}
code. Next, we discuss how we determined the ground truth of
privacy-and-security harming behaviors.  Finally, we demonstrate how we build
our collection of signatures of known-harmful \JS{} behaviors (as determined by
our ground truth), and discuss how we ensured these signatures have high
precision (\IE{} they can accurately detect the same privacy-and-security
harming behaviors occurring in different code units).

\subsection{Difficulties in Building \JS{} Signatures}
\label{sec:meth:difficulties}
Building accurate signatures of \JS{} behavior is difficult for many
reasons, many unique to the browser environment.
First, fingerprinting \JS{} code on the web requires instrumenting both the
\JS{} runtime \textit{and} the browser runtime, to capture the downstream
effects of \JS{} DOM and Web API operations.  For example, \JS{} code can
indirectly trigger a network request by setting the \texttt{src} attribute on an
\texttt{<img>} element.\footnote{Google Analytics, for example, uses this
pattern.} Properly fingerprinting such behavior requires capturing both the
attribute modification and the resulting network request, even though the
network request is not \textit{directly} caused by the script.  Other complex
patterns that require instrumenting the relationship between the \JS{} engine
and the rendering layer include the unpredictable effects of writing to
\texttt{innerHTML}, or writing text inside a \texttt{<script>} element, among
many others.

Second, the web programming model, and the extensive optimizations applied by
\JS{} engines, make attributing script behaviors to code units difficult.
Callback functions, \texttt{eval}, scripts inlined in HTML attributes and \JS{}
URLs, \JS{} microtasks,\footnote{\JSMicrotasksURL{}} and in general the
async nature of most Web APIs make attributing \JS{} execution to its
originating code unit extremely difficult, as described by previous
work.\footnote{Section 2.C. of \cite{iqbal2020adgraph} includes more discussion
of the difficulties of \JS{} attribution} Correctly associating \JS{} behaviors
to the responsible code unit requires careful and extensive instrumentation
across the web platform.

Third, building signatures of \JS{} code on the web is difficult because of the
high amount of indeterminism on the platform.  While in general \JS{}
code runs single threaded, with only one code unit executing at a time,
there is indeterminism in the ordering of events, like network requests starting
and completing, behaviors in other frames on the page, and the interactions
between CSS and the DOM that can happen in the middle of a script executing.
Building accurate signatures for \JS{} behavior on the web requires carefully
dealing with such cases, so that generated signatures include only behaviors and
modifications deterministically caused by the \JS{} code unit.

\subsection{Signature Generation}
\label{sec:meth:gen}

Our solution for building per-event-loop signatures of \JS{} behavior on the web
consists of four parts: %
\begin{enumerate*}[label=(\roman*)]
  \item accurately attributing DOM modifications and Web API accesses to the
    responsible \JS{} unit
  \item enumerating which events occur in a deterministic order (and
      excluding those which vary between page executions)
  \item extracting both immediate and downstream per-event-loop-turn activities
  \item post-processing the extracted signatures to address possible ambiguities.
\end{enumerate*}

This subsection proceeds by giving a high-level overview of each step, enough to
evaluate its correctness and boundaries, but excluding some low-level details we
expect not to be useful for the reader. However, we are releasing all of the
code of this project to allow for reproducibility of our results and further
research~\cite{semanticsignatures}.

\subsubsection{\JS{} Behavior Attribution}
\begin{figure}[t]
  \centering
  \includegraphics[width=\linewidth]{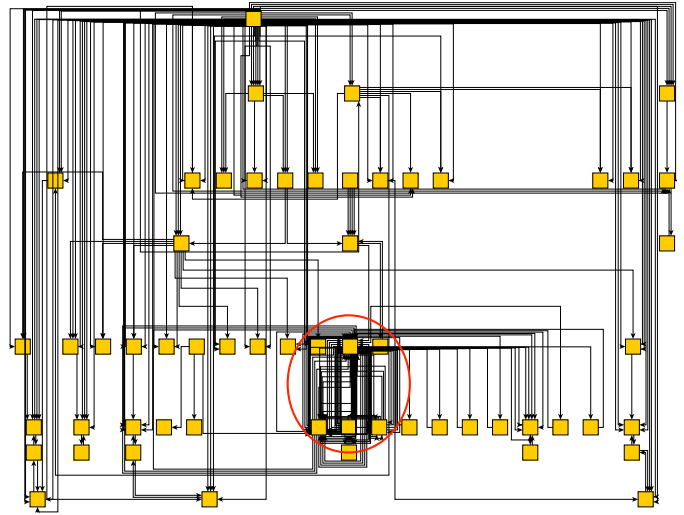}
  \caption{Simplified rendering of execution graph for \url{https://theoatmeal.com}.
    The highlighted section notes the subgraph attributed to Google Analytics tracking
    code.}
  \label{fig:graph-example}
\end{figure}

The first step in our signature-generation pipeline is to attribute
all DOM modifications, network requests, and Web API accesses to the
responsible actor on the page, usually either the parser or a \JS{} unit.
This task is deceptively difficult, for the reasons discussed in Section
\ref{sec:meth:difficulties}, among others.

To solve this problem, we used and extended
\PG{},\footnote{\PageGraphWikiURL{}}
a system for representing the execution of a page as a directed graph.
\PG{} uses nodes to represent elements in a website's environment
(\EG{} DOM nodes, \JS{} units, fetched resources, etc.) and edges to describe
the interaction between page elements. For example, an edge from a script
element to a DOM node might encode the script setting an attribute on that
DOM node, while an edge from a DOM node to a network resource might encode
an image being fetched because of the \texttt{src} attribute on an
\texttt{<img>} node. Figure~\ref{fig:graph-example} provides a
simplified example of a graph generated by \PG{}.

All edges and nodes in the generated graphs are fully ordered, so that
the order that events occurred in can be replayed after the fact.  Edges
and nodes are richly annotated and describe, for example, the type
of DOM node being created (along with parents and siblings it inserted
alongside), the URL being fetched by a network request, or which internal
V8 script id\footnote{\VEightScriptIdURL{}} a code unit in the graph
represents.

We use \PG{} to attribute all page activities to their responsible party.
In the following steps we use this information to determine what each
script did during each turn of the event loop.

\subsubsection{Enumerating Deterministic Script Behaviors}
\begin{table}[tb]
  \begin{threeparttable}
    \begin{tabular*}{\linewidth}{lll}
      \toprule
        Instrumented Event        & Privacy? & Deterministic? \\
      \midrule
        HTML Nodes \\
      \midrule
        \ctab{} node creation           & no & yes \\
        \ctab{} node insertion          & no & yes \\
        \ctab{} node modification       & no & yes \\
        \ctab{} node deletion           & no & yes \\
        \ctab{} remote frame activities & no & no \\
      \midrule
        Network Activity \\
      \midrule
        \ctab{} request start    & yes & some\tnote{1} \\
        \ctab{} request complete & no  & some\tnote{1} \\
        \ctab{} request error    & no  & some\tnote{1} \\
      \midrule
        API Calls \\
      \midrule
        \ctab{} timer registrations          & no          & yes \\
        \ctab{} timer callbacks              & no          & no \\
        \ctab{} \JS{} builtins               & no          & some\tnote{2} \\
        \ctab{} storage access               & yes         & yes\tnote{3} \\
        \ctab{} other Web APIs               & no\tnote{4} & some \\
      \bottomrule
    \end{tabular*}
    \begin{tablenotes}
      \item[1] Non-async scripts, and sync AJAX, occur in a deterministic order.
      \item[2] Most builtins occur in deterministic order (e.g. Date API),
        though there are exceptions (e.g. setTimeout callbacks).
      \item[3] \texttt{document.cookie}, \texttt{localStorage},
        \texttt{sessionStorage}, and \texttt{IndexedDB}
      \item[4] While many Web API can have privacy effects (e.g. WebRTC,
        browser fingerprinting, etc.) we do not consider
        such cases in this work, and focus only on the subset of
        privacy-sensitive behaviors relating to storage and network events.
    \end{tablenotes}
  \end{threeparttable}
  \caption{Partial listing of events included in our signatures, along with
    whether we treat those events as privacy relevant, and whether they occur
    in a deterministic order, given the same \JS{} code.}
  \label{table:signature-events}
\end{table}

Next, we selected page events that will happen in a deterministic
order, given a fixed piece of \JS{} code.  While events like DOM modifications
and calls to (most) \JS{} APIs will happen in the same order each time the same
script is executed, other relevant activities (\EG{} the initiation of most
network requests and responses, timer events, activities across frames) can
happen in a different order each time the same \JS{} code is executed. For our
signatures to match the same \JS{} code across executions, we need to exclude
these non-deterministic behaviors from the signatures that we generate.

Table \ref{table:signature-events} presents a partial listing of which browser
events occur in a deterministic order (and so are useful inputs to code
signatures) and which occur in a non-deterministic ordering (and so should not
be included in signatures).

\subsubsection{Extracting Event-Loop Signatures}
\label{sec:meth:building-sigs}
\begin{figure}[t]
  \centering
  \includegraphics[width=\linewidth]{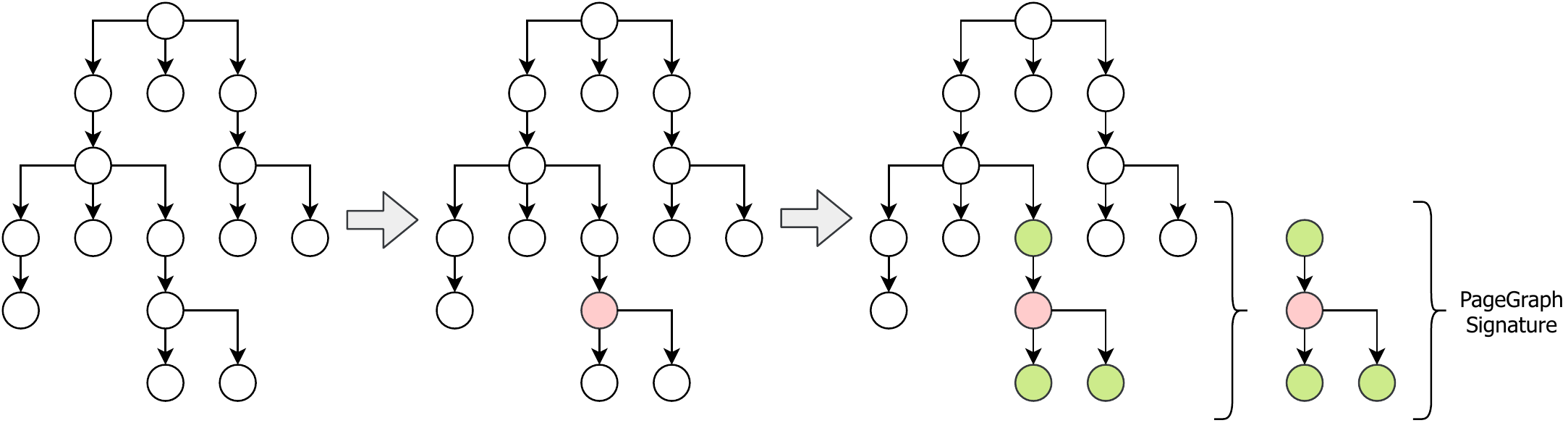}
  \caption{PageGraph signature generation. The red node represents a script unit that executed privacy-related activity and the green nodes are the ones affected by the script unit during one event loop turn. The extracted signature is a \SG{} of the overall \PG{}.}
  \label{fig:signature}
\end{figure}

\begin{figure}[t]
  \centering
  \includegraphics[width=\columnwidth]{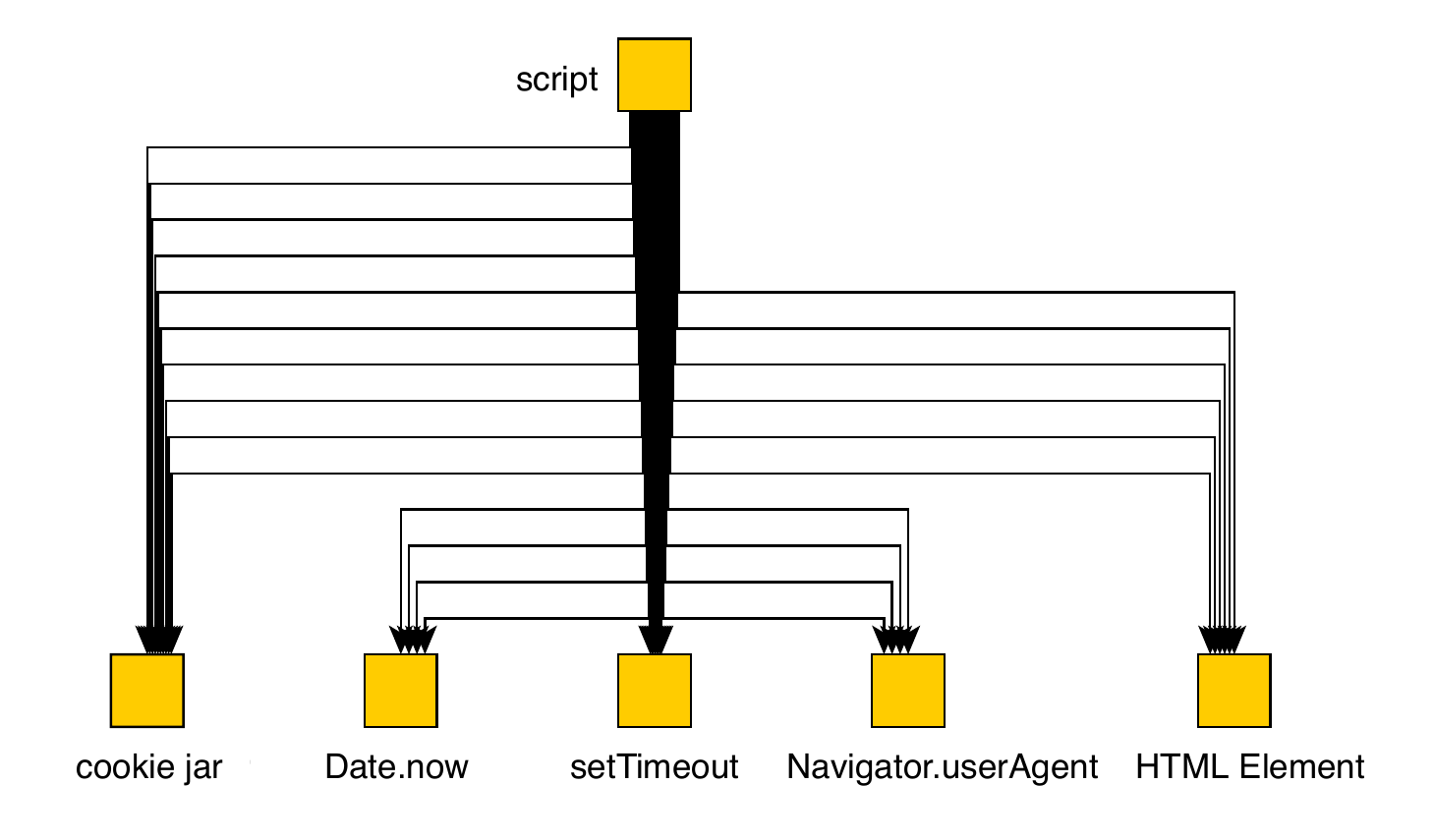}
  \caption{Simplified signature of Google Analytics tracking code.
    Edges are ordered by occurrence during execution, and nodes depict
    Web API and DOM elements interacted with by Google Analytics.}
  \label{fig:google-analytics-sig}
\end{figure}

Next, we use the \PG{} generated graph representation of page execution, along
with the enumeration of deterministic behaviors, to determine the behaviors of
each \JS{} unit during each event loop turn (along with deterministically
occurring ``downstream'' effects in the page). Specifically, to obtain the
signature of \JS{} behaviors that happened during one event-loop turn, our
approach extracts the subgraph depicting the activities of each \JS{} unit, for
each event loop turn, that occurred during page execution (see
Figure~\ref{fig:signature}). However, for the sake of efficiency, we do not
generate signatures of script activity that do not affect privacy. Put
differently, each signature is a \SG{} of the entire \PG{} generated graph, and
\emph{encodes at least one privacy-relevant event}. Hereafter we refer to the
extracted signatures, which depict the per-event-loop behaviors of \JS{}, as
\emph{event-loop signatures}.

For the purposes of this work, we consider privacy-and-security related
events to consist solely of %
\begin{enumerate*}[label=(\roman*)]
  \item storage events,\footnote{i.e. cookies, localStorage,
    sessionStorage, IndexedDB} because of their common use in tracking, and
  \item network events,\footnote{both direct from script (\EG{} AJAX, fetch) and
    indirect (\EG{} \texttt{<img>})} since identifiers need to be exfiltrated
    at some point for tracking to occur.
\end{enumerate*}
We note that there are other types of events that could be considered here,
such as browser fingerprinting-related APIs, but reserve those for future
work.

As an example, Figure~\ref{fig:google-analytics-sig} shows a (simplified)
\SG{} of the larger graph from Figure~\ref{fig:graph-example}, depicting
what the Google Analytics script did during a single event loop turn: accessing
cookies several times (storage events), reading the browser user-agent string,
creating and modifying an \texttt{<img>} element (and thus sending out network
requests), etc.

At a high level, to extract event-loop signatures from a \PG{} generated graph,
we determined which \JS{} operations occurred during the same event-loop turn by
looking for edges with sequential ids in the graph, all attached to or
descending from a single script unit.  As a page executes, control switches
between different script units (or other actions on the page); when one script
yields its turn on the event loop, and another script begins executing, the new
edges in a graph will no longer be attached to the first script, but to the
newly executing one. Event-loop turns are therefore encoded in the graph as
\SG{}s with sequential edges, all related to the same script node.  We discuss
some limitations of this approach, and why we nevertheless preferred it to
possible alternatives in Section\ref{sec:discussion:limitiations}.

More formally, we build signatures of privacy-and-security affecting \JS{}
behavior using the following algorithm:\footnote{This description omits some
implementation specific details and post-processing techniques that are
not fundamental to the approach. They are fully documented and described in
our shared source code~\cite{semanticsignatures}}.

\begin{enumerate}[label=(\roman*)]
  \item Extract all edges in the graph representing a privacy-effecting \JS{}
    operation (as noted in Table~\ref{table:signature-events}).
  \item Attribute each of these edges to the \JS{} unit responsible.  If no
    script is responsible (\EG{} a network request was induced by the parser),
    abort.
  \item Extract the maximum \SG{} containing the relevant edge and responsible
    \JS{} code unit comprising all sequentially occurring nodes and edges. This
    is achieved by looking for edges that neighbor the \SG{}, and which occurred
    immediately before (or after) the earliest (or latest) occurring event
    in the \SG{}. If an edge is found, add it and the attached nodes to
    the \SG{}.
  \item Repeat step 3 until no more edges can be added to the \SG{}.
\end{enumerate}

Once each \SG{} is extracted, a hash representation is generated by removing
any edges that represent non-deterministically ordered events (again, see Table
\ref{table:signature-events}), chronologically ordering the remaining edges and
nodes, concatenating each elements' type (but omitting other attributes), and
hashing the resulting value.  This process yields a SHA-256 signature for the
deterministic behavior of every event-loop turn during which a \JS{} unit
carried out at least one privacy-relevant operation.

\subsection{Privacy Behavior Ground Truth}
\label{sec:meth:truth}
Next, we need a ground truth set of privacy harming signatures, to build
a set of known privacy-harming \JS{} behaviors.  We then use this ground
truth set of signatures to look for instances where the same privacy-harming
code reoccurred in \JS{} code not blocked by current content blockers,
and thus evaded detection.

We used \EL{} and \EP{} to build a ground truth determination of privacy-harming
\JS{} behaviors.  If a script was identified by an \EL{} or \EP{} filter rule
for blocking, and was not excepted by another rule, then we considered all the
signatures generated from that code as privacy-harming, and thus should be
blocked. This measure builds on the intuition that filter rules block known bad
behavior, but miss a great deal of additional unwanted behavior (for the reasons
described in Section \ref{sec:problem}). Put differently, this approach models
filter lists as targeting behaviors in code units (and that they target URLs as
an implementation restriction), and implicitly assumes that filter lists have
high precision but low (or, possibly just lower) recall in identifying
privacy-and-security harming behaviors.

To reduce the number of \JS{} behaviors falsely labeled as privacy harming, we
removed a small number of filter list network rules that blocked all script on a
known-malware domain. This type of rule does not target malicious or unwanted
resources, but \textit{all} the resources (advertising and tracking related, or
otherwise) fetched by the domain. As these rules end up blocking malicious and
benign resources alike, we excluded them from this work. An example of such a
rule is \texttt{\$script,domain=imx.to}, taken from \EL{}.

\subsection{Determining Privacy-Harming Signatures}
\label{sec:meth:privacy-sigs}

To generate a collection of signatures of privacy-harming \JS{} behaviors on the
web, we combine our algorithm that extracts event-loop signatures
(Section~\ref{sec:meth:gen}), with the ground truth of privacy-harming
behaviors given by EL/EP (Section~\ref{sec:meth:truth}). Specifically, we
produce this collection of signatures by visiting the Alexa top 100K websites
and recording their graph representations (one graph per visited website), using
our PageGraph-enhanced browser. For each visited website, we gather from its
graph representation every script unit executing on the page, including remote
scripts, inline scripts, script executing as \JS{} URLs, and scripts defined in
HTML attributes. We then extracted signatures of \JS{} behavior during each
event loop turn, and recorded any scripts that engaged in privacy-relevant
behaviors.

Next, we omitted signatures that were too small to be highly identifying
from further consideration. After an iterative process of sampling and manually
evaluating code bodies with signature matches, we decided to only consider
signatures that consisted of at least \SignatureMinEdges{} \JS{} actions
(encoded as \SignatureMinEdges{} edges), and interacting with at least
\SignatureMinNodes{} page elements (encoded as nodes, each representing a DOM
element, \JS{} builtin or privacy-relevant Web API endpoint).

This minimal signature size was determined by starting with an initial signature
size of 5 edges and 4 nodes, and then doing a manual evaluation of 25 randomly
sampled matches between signatures (\IE{} cases where the same signature
was generated by a blocked and not-blocked script). We had our domain-expert
then examine each of the 25 randomly sampled domains to determine whether the
code units actually included the same code and functionality. If the expert
encountered a false positive (\IE{} the same signature was generated by code
that was by best judgement unrelated) the minimum graph size was increased, and
25 new matches were sampled.  This process of manual evaluation was repeated
until the expert did not find any false positives in the sampled matches,
resulting in a minimum graph size of \SignatureMinEdges{} edges and
\SignatureMinNodes{} nodes.

Finally, for scripts that came from a URL, we noted whether the script was
associated with advertising and/or tracking, as determined by \EL{} and \EP{}.
We labeled all signatures generated by known tracking or advertising scripts as
privacy-harming (and so should be blocked by a robust content blocking tool).
We treated signatures from scripts not identified by \EL{} or \EP{}, but which
matched a signature from a script identified \EL{} or \EP{}, as \textit{also}
being privacy-harming, and so evidence of filter list evasion.  The remaining
signatures (those from non-blocked scripts, that did not match a signature from
a blocked script) were treated as benign.  The results of this measurement
are described in more detail in Section \ref{sec:results}.

\section{Results}
\label{sec:results}

In this section we report the details of our web-scale measurement of filter
list evasion, generated by applying the techniques described in
Section~\ref{sec:meth} to the Alexa 100K.  The section proceeds by
first describing the raw website data gathered during our crawl, then discusses
the number and size of signatures extracted from the crawl. The section
follows with measurements of how this evasion impacts browsing
(\IE{} how often users encounter privacy-and-security harming behaviors that
are evading filter lists) and concludes with measurements of what web
parties engage in filter list evasion.

\subsection{Initial Web Crawl Data}
\begin{table}[tb]
  \centering
  \begin{tabular}{lr}
    \toprule
      Measurement & Value \\
    \midrule
      Crawl starting date         & \CrawlStartDate{} \\
      Crawl ending date           & \CrawlEndDate{} \\
      Date of filter lists        & \CrawlFilterListDate{} \\
      Num domains crawled          & 100,000 \\
      Num domains responded        & \CrawlRespondingDomains{} \\
      Num domains recorded         & \CrawlPageGraphDomains{} \\
    \bottomrule \\
  \end{tabular}
  \caption{Statistics regarding our crawl of the Alexa 100k, to both build
    signatures of known tracking code, and to use those signatures to
    identify new tracking code.}
  \label{table:crawl-statistics}
\end{table}


We began by using our PageGraph-enhanced browser to crawl the Alexa 100K,
which we treated as representative of the web as a whole.  We automated
our crawl using a puppeteer-based tool, along with extensions to \PG{} to
support the DevTools
interface~\footnote{\url{https://chromedevtools.github.io/devtools-protocol/}}.

For each website in the Alexa 100K, our automated crawler visited the domain's
landing page and rested for 60 seconds to allow for sufficient time for scripts
on the page to execute. We then retrieved the \PG{} generated
graph-representation of each page's execution, encoded as a GraphML-format XML
file.

Table~\ref{table:crawl-statistics} presents the results of this crawl.  From the
Alexa 100K, we got a successful response from the server
from \CrawlRespondingDomains{} domains, and were able to generate the graph
representation for \CrawlPageGraphDomains{}.  We attribute not being able
successfully crawl \CrawlNotRespondingDomains{} domains to a variety of factors,
including bot detection scripts~\cite{invernizzi2016cloak}, certain sites being
only accessible from some IPs~\cite{tschantz2018bestiary, afroz2018exploring},
and regular changes in website availability among relatively unpopular domains.
This number of unreachable domains is similar to those found by other automated
crawl studies~\cite{snyder2016browser, jueckstock2019blind}. A further
\CrawlNonPageGraphDomains{} domains could not be measured because they used
browser features that \PG{} currently does not correctly attribute (most
significantly, module scripts).

\subsection{Signature Extraction Results}
\label{sec:results:overview}
\begin{table*}[tb]
  \centering
  \begin{tabular}{lrr}
    \toprule
       & \vtop{\hbox{\strut\# Scripts Matched by}\hbox{\strut Ground Truth Signatures}}
      & \vtop{\hbox{\strut \# Scripts Matched}\hbox{\strut by Small Signatures}}\\
    \midrule
      Scripts generating relevant signatures (unique) & \NumFingerprintedScriptsUnique{} & \NumSmallFingerprintedScriptsUnique{} \\
      Scripts blocked by EL/EP (total) & \NumBlockedFingerprintedScriptsInstances{} & \NumSmallBlockedFingerprintedScriptsInstances{} \\
      Scripts blocked by EL/EP (unique) & \NumBlockedFingerprintedScriptsUnique{} & \NumSmallBlockedFingerprintedScriptsUnique{} \\
      External scripts not blocked (total) & \NumNewFingerprintedScriptURLInstances{} & \NumSmallNewFingerprintedScriptURLInstances{} \\
      External scripts not blocked (unique) & \NumNewFingerprintedScriptURLs{} & \NumSmallNewFingerprintedScriptURLs{} \\
      Inline scripts not blocked & \NumEvasionByInlining{} & \NumSmallEvasionByInlining{} \\
      Total unique scripts not blocked (external + inline) & \NumNewFingerprintedScriptsUnique{} & \NumSmallNewFingerprintedScriptsUnique{} \\
    \bottomrule \\
  \end{tabular}
  \caption{The number of scripts whose behaviors match signatures from our ground truth set, both in total and broken down by whether they are blocked by EL/EP. For comparison we also show the same statistics for the discarded small signatures.}
  \label{table:signature-stats}
\end{table*}

Next, we run our signature generation algorithm (Section~\ref{sec:meth:gen}) on
the graph representation of the \CrawlPageGraphDomains{} websites that we
crawled successfully from the Alexa top 100K. In total this yielded
\NumSignatures{} signatures from all the encountered scripts (of these
\NumSignatures{} generated signatures, \NumSignaturesUnique{} are unique; the
same script can be included in multiple websites and thus generate the same
signatures for those websites). We then filtered this set of signatures to those
matching the following criteria:

\begin{enumerate}
  \item Contained at least one privacy-or-security relevant event (defined in
    Section~\ref{sec:meth:building-sigs})
  \item Occurred at least once in a script blocked by \EL{} or \EP{} (\IE{} a
    \emph{blocked script})
  \item Occurred at least once in a script \textit{not} blocked by \EL{} and
    \EP{} (\IE{} an \emph{evaded script}).
  \item Have a minimum size of \SignatureMinEdges{} edges and
    \SignatureMinNodes{} nodes (see Section~\ref{sec:meth:privacy-sigs})
\end{enumerate}

This filtering resulted in \NumBlockedSignatures{} \emph{unique} signatures. We
refer to this set of signatures as \emph{ground truth signatures}. Our goal here
is to focus only on the signatures of behaviors that are identified by \EL{} and
\EP{} as privacy-harming, but also occur in other scripts not blocked by these
filter lists.  Recall from Section~\ref{sec:meth:privacy-sigs} that we impose a
lower bound (\SignatureMinEdges{} edges and \SignatureMinNodes{} nodes) on the
signature size determined manually by our domain expert in order to reduce false
positives. If we remove the restriction on the minimum signature size, then the
above filtering would give us a total of \NumBlockedAndSmallSignatures{} unique
signatures (\IE{} \NumSmallSignatures{} were discarded as too small).

Table~\ref{table:signature-stats} summarizes the scripts from which our
signature generation algorithm (Section~\ref{sec:meth:gen}) produced at least
one signature in our ground truth set, both in total and broken down according
to whether they are blocked by \EL{} and \EP{}. For comparison, we also show the
corresponding statistics for the \NumSmallSignatures{} signatures that we
discarded as too small. Not surprisingly, the discarded small signatures were
found in more scripts than our ground truth set. This is because the specificity
of a signature is proportional to the number of script actions that it registers
(\EG{} a signature consisting of only one storage write operation would be found
in many scripts that use the local storage API).

For our purposes we prefer precision over recall, by utilizing
expert domain knowledge to set a reasonable cut-off signature size. Notice that
our approach is optimized towards minimizing false positives, which means that
the behavior of the script needs to be expressive enough (have enough
edges/nodes) to indicate privacy-harming behavior (see
\S\ref{sec:meth:privacy-sigs}). Small signatures are less expressive, so they
resulted in our experiments in matching more scripts, which include both
true/false positives.

\begin{figure}[t]
  \centering
  \includegraphics[width=\columnwidth]{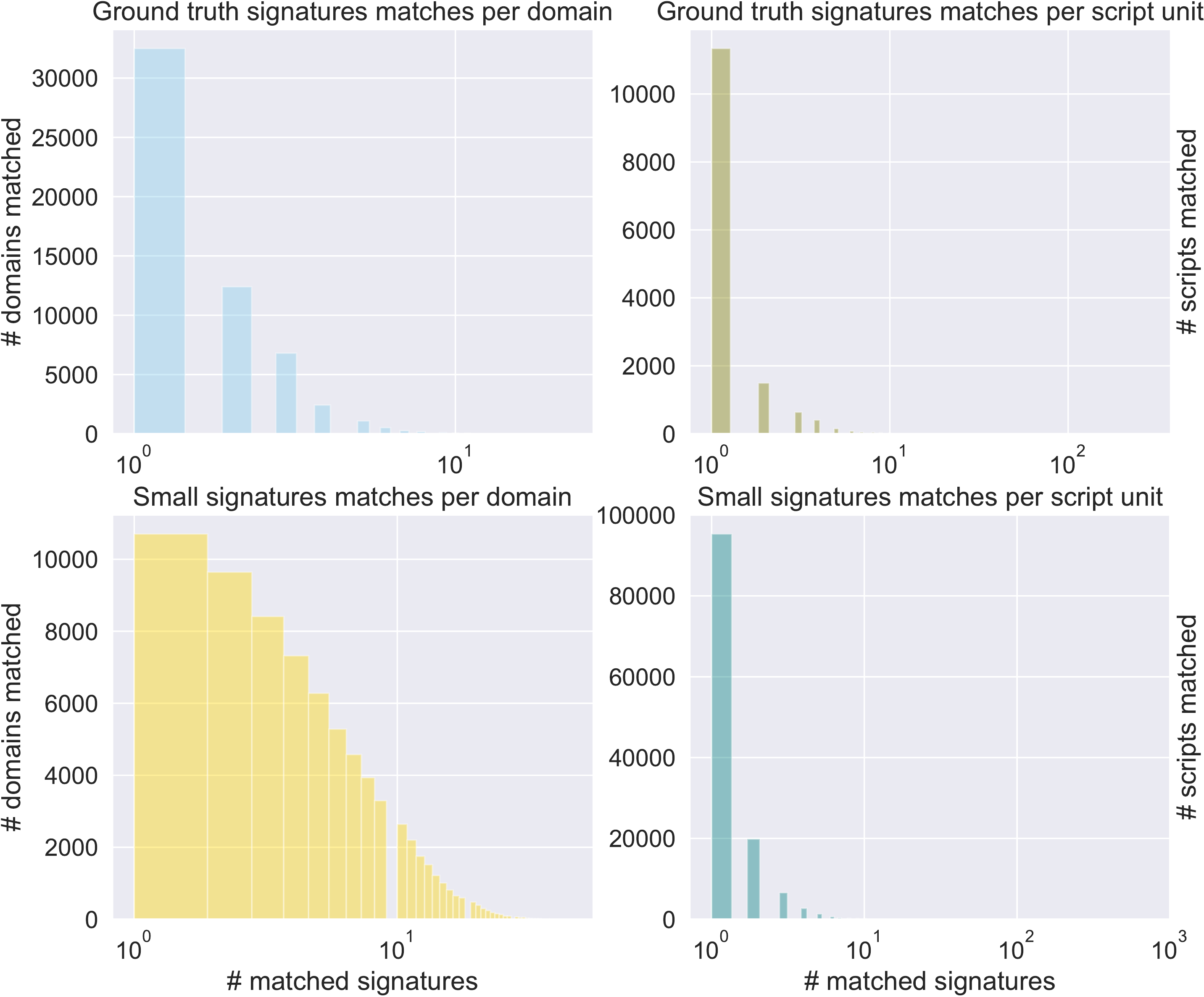}
  \caption{Distribution of the number of signatures per domain and the number of such signatures in each matched script unit for our ground truth dataset and for the small signatures dataset.}
  \label{fig:sigs-per-website-script}
\end{figure}

Figure~\ref{fig:sigs-per-website-script} shows the distribution of the number of
unique signatures in our ground truth set that were found in scripts on each
visited domain in our crawl of the Alexa top 100K (\NumDomainsMatchGroundTruthSigs{} domains
have at least one script where signatures from our ground truth set were found), as well as
the distribution of the number of unique ground truth signatures in each script
unit where such signatures were found. As we did in Table~\ref{table:signature-stats}, for comparison here we also plot the same statistics for the small signatures.

In total, our ground truth signatures identified
\NumNewFingerprintedScriptURLs{} new unique external script URLs
(\NumNewFingerprintedScriptURLInstances{} instances) hosting known-harmful
behavior, but missed by filter lists, an increase in
\PctMoreNewFingerprintedScriptURLs{} identified harmful URLs (when measured
against the number of scripts only identified by filter lists and which contain
the ground truth signatures). These evading scripts were hosted on
\NumDomainsHostingFingerprintedScripts{} unique domains. In addition to these
evaded external scripts, our signatures also matched \emph{inline} scripts.
Inline scripts are those whose \JS{}
source is contained entirely within the text content of a \texttt{script} tag,
as opposed to external scripts whose URL is encoded in the \texttt{src}
attribute of a \texttt{script} tag, and thus cannot be  blocked by existing
tools. We identified \NumEvasionByInlining{} instances of privacy-relevant
behavior from EL/EP blocked scripts moved inline, carried out on
\NumWebsitesEvasionByInlining{} domains.

\subsection{Impact on Browsing}
\begin{figure}[t]
  \centering
  \includegraphics[width=.8\columnwidth]{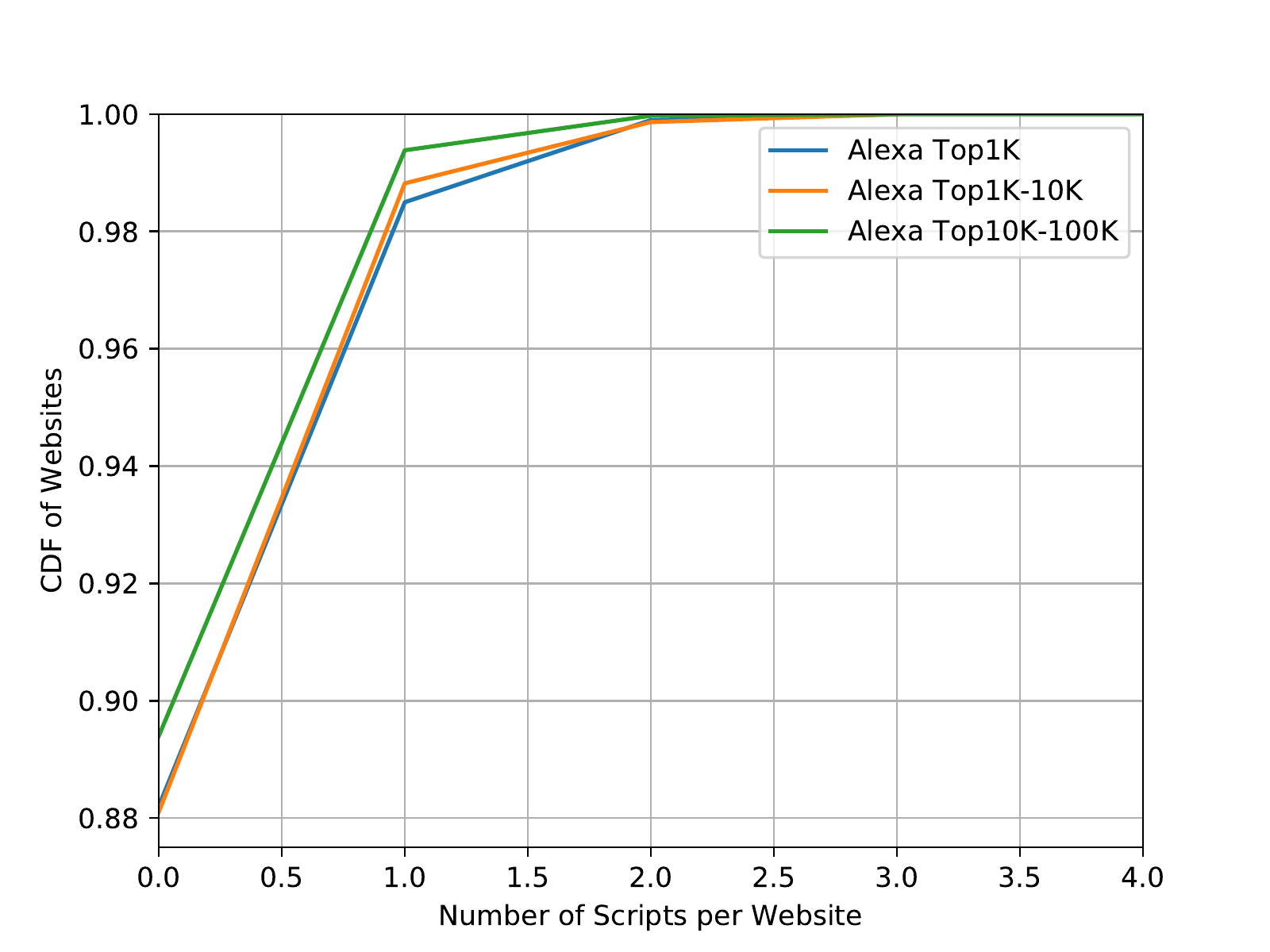}
  \caption{Total number of evaded scripts per website, for ``popular'' (Alexa top 1K), ``medium'' (Alexa top 1K - 10K), and ``unpopular'' (Alexa top 10K - 100K) websites.}
  \label{fig:alexa-1k-evasion}
\end{figure}

Next, we attempt to quantify the practical impact on privacy and security
from filter list evasion.  Here the focus is not on the number of parties or
scripts engaging in filter list evasion, but on the number of websites
users encounter on which filter list evasion occurs.  We determined this
by looking for the number of sites in the Alexa 100K (in the subset we
were able to record correctly) that included at least one script matching
a signature from a blocked script, but which was not blocked.

We find that \NumWebsitesWithEvasion{} of the \CrawlPageGraphDomains{}
domains measured included at least one known privacy-or-security harming
behavior that was not blocked because of filter list evasion.  Put differently,
\PctWebsitesWithEvasion{} of websites include at least one instance of
known-harmful functionality evading filter lists.

We further measured whether these evasions occurred more frequently on
popular or unpopular domains.  We did so by breaking up our data set
into three groups, and comparing how often filter list evasion occurred
in each set.  We divided our dataset as follows:

\begin{enumerate}
  \item \textbf{Popular sites}: Alexa rank 1--1k
  \item \textbf{Medium sites}: Alexa rank 1,001--10k
  \item \textbf{Unpopular sites}: Alexa rank 10,001--100k
\end{enumerate}

Figure~\ref{fig:alexa-1k-evasion} summarizes this measurement as a CDF
of how many instances of filter list evasion occur on sites in each group.
As the figure shows, filter list evasion occurred roughly evenly on sites
in each group; we did not observe any strong relationship between site
popularity and how frequently filter lists were evaded.

\subsection{Popularity of Evading Parties}

Finally, we measured the relationship, in regards to domain popularity, between
the sites originally hosting the scripts blocked by \EL{} and \EP{}, and the
sites that host scripts with the same privacy-harming behaviors but evade
filter list blocking. Our goal in this measurement is to understand if
harmful scripts are being moved from popular domains (where they are more likely
to be encountered and identified by contributors to crowdsourced filter lists)
to less popular domains (where the domain can be rotated frequently).
Specifically, we want to calculate the delta of the Alexa rank for where
code was blocked, and code with the same behavior was not blocked.

We determine these ranking deltas by extracting from our results all the
\emph{unique} pairs of domains that host scripts matching the same signature
of privacy-affecting behavior. That is, for a given signature in our ground
truth signature set, if there are \texttt{n} unique domains hosting blocked
scripts matching that signature, and
\texttt{m} unique domains hosting evading scripts matching the same signature,
then we would have \(\texttt{n} \times \texttt{m}\) domain pairs for that
signature.

We arrange the domains in each pair as a tuple \texttt{(from\_domain,
to\_domain)} to signify the fact that the scripts hosted on the
\texttt{to\_domain} contain the same privacy-harming semantics as those on the
\texttt{from\_domain}, and that the scripts hosted on the \texttt{to\_domain}
are not blocked by filter lists. Note that the final set of domain
pairs that we extract across all ground truth signatures contain only unique
pairs (\EG{} if the domain pair \texttt{(s, t)} is extracted for both signature
\texttt{sig1} and \texttt{sig2}, then \texttt{(s, t)} appears only once in the
final set of domain pairs).

In total we collected 9,957 such domain pairs. For the domains in each
pair, we then look up their Alexa rankings and calculate their delta as the
ranking of \texttt{from\_domain} subtracted by \texttt{to\_domain} (\IE{} a
negative delta means \texttt{to\_domain} is less popular than
\texttt{from\_domain}). We note that at present since we only have the rankings
for Alexa top one million domains, there are 2,898 domain pairs which we do not
have the ranking information for either the \texttt{from\_domain} or the
\texttt{to\_domain} (\IE{} their popularity ranking is outside of the top one
million). We use a ranking of one million whenever we cannot determine the
ranking of a domain.

\begin{figure}[t]
  \centering
  \includegraphics[width=.8\columnwidth]{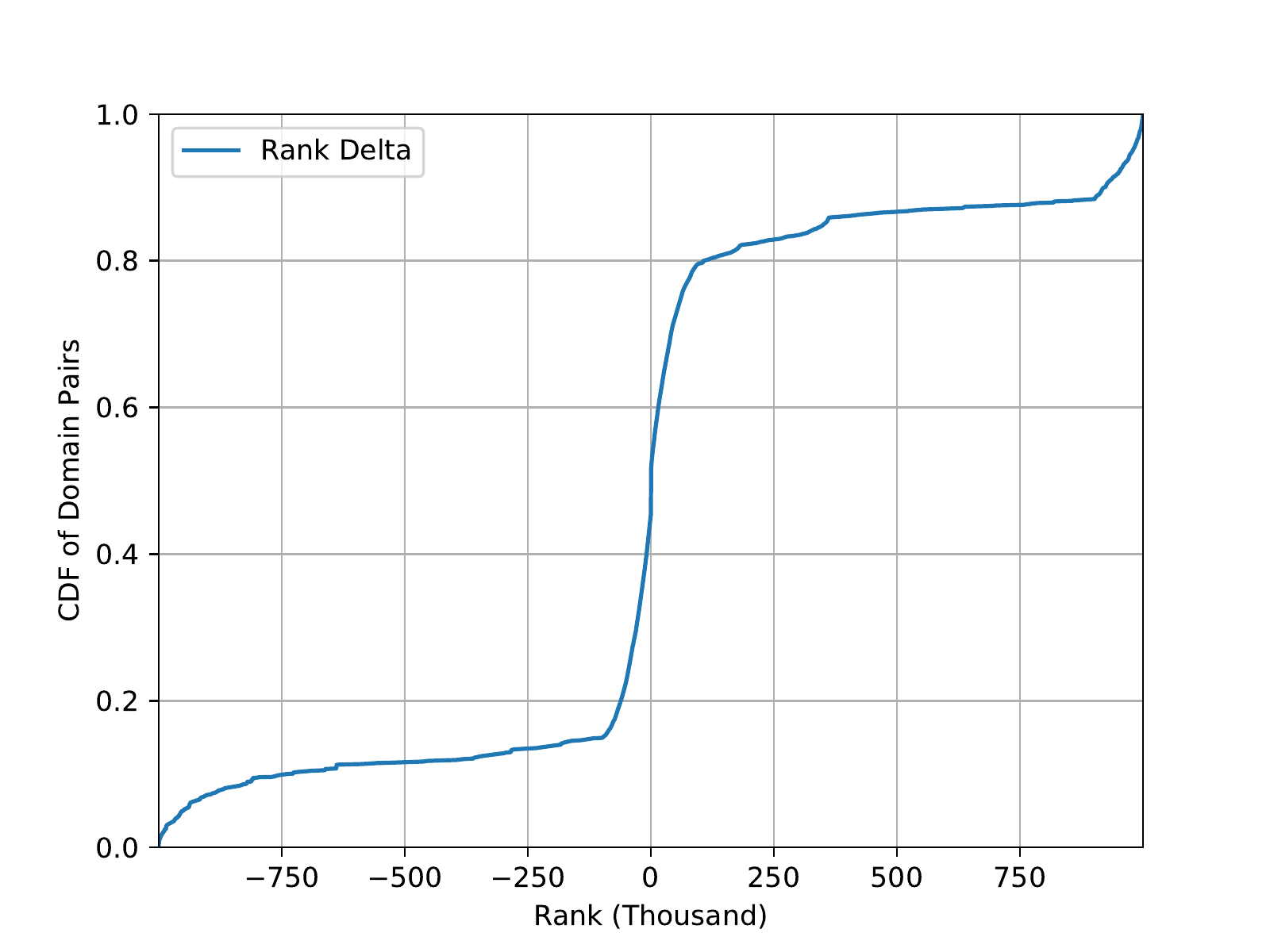}
  \caption{Distribution of the delta in Alexa ranking of domains hosting EL/EP blocked scripts vs. evaded scripts that matched the same signature. A negative delta means the script is moved from a popular domain to a less popular domain. The x-axis of domain rank delta is in thousands.}
  \label{fig:evasion-rank-deltas}
\end{figure}

Figure~\ref{fig:evasion-rank-deltas} shows the distribution of all the ranking
deltas (all pairs). Note that this distribution is very symmetric: about as many
of the domain pairs have negative delta as those that have positive delta, and the
distribution on both sides of the x-axis closely mirrors each other. We believe
this is mostly due Alexa favoring first-party domains when calculating
popularity-metrics; according to Alexa's documentation, multiple requests to the same URL by
the same user counts as only one page view for that URL on that day~\cite{how_alexa_ranking}.
Thus, if on a single day the user visits multiple sites that contain tracking
scripts loaded from the same tracker domain, then Alexa counts those as only one
visit to that tracker domain. As a result, domains that host tracking scripts
tend to occupy the middle range of the Alexa ranking, and their tracking scripts
are equally likely to be hosted on websites both before and after them in the
Alexa rankings (web developers often choose host third-party tracking scripts on
domains that they control, while at the same time minifying/obfuscating, or
otherwise bundling the scripts, in order to evade filter list blocking, and we
provide a taxonomy of such evasion attempts in Section~\ref{sec:taxonomy}).

\begin{table}[t]
  \centering
  \begin{tabular}{lrrr}
    \toprule
      Hosting Domain & Requesting Domains & Script URLs & Matches \\
    \midrule
      google-analytics.com & 47,366 & 44 & 55,980 \\
      googletagmanager.com & 6,963 & 6,158 & 6,967 \\
      googlesyndication.com & 5,711 & 38 & 5,711 \\
      addthis.com & 1,600 & 51 & 2,464 \\
      facebook.net & 1,479 & 1,313 & 1,479 \\
      adobedtm.com & 1,076 & 1,133 & 2,973 \\
      amazon-adsystem.com & 915 & 1 & 943 \\
      adroll.com & 814 & 5 & 1,931 \\
      doubleclick.net & 774 & 5 & 985 \\
      yandex.ru & 610 & 3 & 684 \\
    \bottomrule \\
  \end{tabular}
  \caption{Most popular domains hosting resources that were blocked by filter lists.
  The first column records the hosting domain, the next column the number of domains
  loading resources from the hosting domain, the third column the number of unique
  URLs requested from the domain, and the final column the count of (non-unique) blocked, harmful scripts loaded from the domain.}
  \label{table:top-blocked-domains}
\end{table}
\begin{table}[t]
  \centering
  \begin{tabular}{lrrr}
    \toprule
      Hosting Domain & Requesting Domains & Script URLs & Matches \\
    \midrule
      google-analytics.com & 5,157 & 4 & 6,412 \\
      addthis.com & 1,596 & 50 & 2,455 \\
      shopify.com & 543 & 4 & 545 \\
      adobedtm.com & 398 & 331 & 756 \\
      tiqcdn.com & 311 & 248 & 709 \\
      googletagservices.com & 136 & 1 & 143 \\
      segment.com & 114 & 107 & 122 \\
      tawk.to & 85 & 85 & 90 \\
      outbrain.com & 73 & 4 & 78 \\
      wistia.com & 71 & 5 & 85 \\
    \bottomrule \\
  \end{tabular}
  \caption{Most popular domains hosting scripts that evaded filter lists, but matched known
    harmful scripts. The first column records the hosting domain, the second
    column the number of domains that referenced the hosting domain, the third
    column the number of unique, evading urls on the hosting domain, and the
    final column the count of (non-unique) non-blocked, harmful scripts loaded from the domain.}
  \label{table:top-evading-domains}
\end{table}

Additionally, we measured what domains hosted scripts most often blocked by
filter lists, and which domains hosted scripts that contained known-harmful
behavior, but evaded detection.  Tables~\ref{table:top-blocked-domains}
and~\ref{table:top-evading-domains} record the results of these measurements. We
find that Google properties are the most frequently blocked resources on the web
(Table~\ref{table:top-blocked-domains}), both for tracking and advertising
resources, followed by the \texttt{addthis.com} widget for social sharing (that
also conducts tracking operations).

Unsurprisingly then, we also find that these scripts are also the most common
culprits in filter list evasion.  Code originally hosted by Google and AddThis
are the most frequently modified, inlined, moved or bundled to evade filter
list detection.

\section{Evasion Taxonomy}
\label{sec:taxonomy}
\begin{table}[ht]
  \centering
  \begin{tabular}{lrr}
    \toprule
      Technique & \# Instances (\% Total) & Unique Scripts (\% Total) \\
    \midrule
      Moving & \NumInstanceEvasionByMovedScript{} (\PctInstanceEvasionByMovedScript{}) & \NumUniqeScriptsEvasionByMovedScript{} (\PctUniqueEvasionByMovedScript{}) \\
      Inlining & \NumEvasionByInlining{} (\PctInstanceEvasionByInlining{}) & \NumEvasionByInlining{} ( \PctUniqueEvasionByInlining{}) \\
      Bundling & \NumInstanceEvasionByBundling{} (\PctInstanceEvasionByBundling{}) & \NumEvasionByBundling{} (\PctUniqueEvasionByBundling{}) \\
      Common Code & \NumInstanceEvasionByCommonCode{} (\PctInstanceEvasionByCommonCode{}) & \NumEvasionByCommonCode{} (\PctUniqueEvasionByCommonCode{}) \\
    \bottomrule \\
  \end{tabular}
  \caption{Taxonomy and quantification of observed filter list evasion techniques in the Alexa 100k.}
  \label{table:evasion-taxonomy}
\end{table}

This section presents a taxonomy of techniques site authors use to evade filter
lists.  Each involves attackers leveraging the common weakness of current web
content blocking tools (\IE{} targeting well known URLs) to evade defenses and
deliver known privacy-or-security harming behavior to websites.

We observed four ways privacy-and-security harming \JS{} behaviors evade
filter lists: %
\begin{enumerate*}[label=(\roman*)]
  \item moving code from a URL associated with tracking, to a new URL,
  \item inlining code on the page,
  \item combining malicious and benign code in the same file
  \item the same privacy-affecting library, or code subset, being used in two
    different programs.
\end{enumerate*}

Each of the following four subsections defines an item in our
taxonomy, gives a representative observed case study
demonstrating the evasion technique, and finally describes the methodology
for programmatically identifying instances of the evasion technique.
Table \ref{table:evasion-taxonomy} presents the results of applying
our taxonomy to the \NumNewFingerprintedScriptsUnique{} unique scripts
(\NumNewFingerprintedScriptsInstances{} instances) that we identified in
Section~\ref{sec:results:overview} as evading filter lists.

For each taxonomy label, we perform code analysis and comparison techniques
using Esprima,\footnote{\url{https://esprima.org/}} a popular and open-source
\JS{} parsing tool. We use Esprima to generate ASTs for each \JS{} file to look
for structural similarities between code units. By comparing the AST node types
between scripts we are resilient to code modifications that do not affect the
structure of the program, like renaming variables or adding/changing comments.
We consider signatures from scripts \textit{not blocked} by \EL{} and \EP{}, but
matching a signature generated by a script blocked by \EL{} and \EP{} to
determine the relationship of the non-blocked script to the blocked scripts.

Finally, this taxonomy is not meant to categorize or imply the \textit{goals} of
the code or site authors, only the mechanisms that causes the bypassing of
URL-based privacy tools. Additionally, each of the case studies are
current as of this writing.  However, we have submitted fixes and new filter
lists rules to the maintainers of \EL{} and \EP{} to address these cases.  As a
result, sites may have changed their behavior since this was written.

\subsection{Moving Code}
\label{sec:taxonomy:methodology}

The simplest filter list evasion strategy we observed is moving tracking code
from a URL identified by filter lists to a URL unknown to
filter lists. This may take
the form of just copying the code to a new domain but leaving the path
fixed,\footnote{\url{https://tracker.com/track.js} $\rightarrow$
\url{https://example.org/track.js}} leaving the script contents constant but
changing the path,\footnote{\url{https://tracker.com/track.js} $\rightarrow$
\url{https://tracker.com/abdcjd.js}} or some combination of both. We also
include in this category cases where code was moved to a new URL and minified
or otherwise transformed without modifying the code's AST.

Site authors may move code from well-known URLs to unique ones for a variety
of reasons.  In some cases this may be unrelated to evading filter lists.
Changes to company policies might require all code to be hosted on the first
party, for security or integrity purposes.  Similarly, site authors might move
tracking code from the ``common'' URL to a new URL out of some performance
benefit (\EG{} the new host being one that might reach a targeted user
base more quickly).

Nevertheless, site authors also move code to new URLs to avoid filter list rules.
It is relatively easy for filter list maintainers to identify tracking code
served from a single, well known URL, and fetched from popular sites. It is
much more difficult for filter list maintainers to block the
same tracking code served from multitudes of different URLs.

\subsubsection{Classification Methodology}
\label{sec:ast_hash}

We detect cases of ``moving code'' evasion by looking for cases where code with
the identical AST appears at both blocked and not-blocked URLs. For each script
that generated a signature that was \emph{not} blocked by \EL{} or \EP{} (i.e,
an evading script), we first generated the AST of the script, and then generated
a hash from the ordered node types in the AST. We then compared this AST hash
with the AST hash of each blocked script that also produced the same signature.
For any not-blocked script whose AST hash matched one of the blocked scripts, we
labeled that as a case of evasion by ``moving code''. We observed
\NumUniqeScriptsEvasionByMovedScript{} unique script units
(\NumInstanceEvasionByMovedScript{} instances) that evade filter lists using
this technique in in the Alexa 100K.

\subsubsection{Case Study: Google Analytics}
Google Analytics (GA) is a popular tracking (or analytics) script, maintained by
Google and referenced by an enormous number of websites.  Generally websites get
the GA code by fetching one of a small number of well known URLs (\EG{}
\url{https://www.google-analytics.com/analytics.js}). As this code has clear
implications for user privacy, the \EP{} filter list blocks this resource,
with the rule \texttt{||google-analytics.com/analytics.js}.

However, many sites would like to take advantage of GA's tracking
capabilities, despite users' efforts to protect themselves.  From our results,
we see \NumGAEvasionsUnique{} unique cases (\IE{} unique URLs serving the evaded GA code) where site
authors copy the GA code from the Google hosted location and move it to a new,
unique URL. We encountered these \NumGAEvasionsUnique{} new, unique
Google-Analytics-serving URLs on \NumGAEvasionInstances{} sites in the Alexa 100k.
Google Analytics alone comprised \PctGAEvasionsUnique{} and
\PctGAEvasionInstances{} of the unique scripts and instances, respectively, of
all cases in our ``moving code'' category. Most memorably, we found
the GA library, slightly out of date, being served from
\url{https://messari.io/js/wutangtrack.js}, and referenced from \url{messari.io}.

\subsection{Inlining Code}
Trackers and site authors also bypass filter lists by ``inlining'' their
code.  While usually sites reference \JS{} at a URL (\EG{} \texttt{<script src=…>}),
HTML also allows sites to include \JS{} as text in the page
(e.g. \texttt{<script>(code)</script>}, which causes to the browser to execute
script without a network request.

A side effect of this ``inlining'' is that URL-based privacy tools
lack an opportunity to prevent the script's execution.  We note that there
are also additional harms from this practice, most notably performance
(\EG{} code delivered inline is generally not cached, even if its reused on
subsequent page views).  Depending on implementation, inlining scripts
can also delay rendering the page, in a way remote async scripts do not.

\subsubsection{Classification Methodology}
Inlining is the most straightforward evasion type in
our taxonomy scheme. Since \PG{} records the source location of each
\JS{} unit that executes during the loading of a page, we can easily determine
which scripts were delivered as inline code.  We then compare the AST hashes
(whose calculation method we described in Section~\ref{sec:ast_hash})
of all inline scripts to all blocked scripts that generated identical signatures.
We labeled all cases where the hashed-AST of an inline script matched the
hashed-AST of a script blocked by \EL{} or \EP{}, and both scripts generated
identical signatures, as cases of evasion by ``inlining''. We observed \NumEvasionByInlining{} cases of
sites moving blocked, privacy harming behaviors inline.

\subsubsection{Case Study: Dynatrace}
Dynatrace is a popular \JS{} analytics library that allows site owners to
monitor how their web application performs, and to learn about their users
behavior and browsing history.  It is typically loaded as a third-party
library by websites, and is blocked in \EP{} by the filter rule
\texttt{||dynatrace.com$\hat{~}$\$third-party}.  Similar, client-specific versions of the
library are also made available for customers to deploy on their domains, which
\EP{} blocks with the rule \texttt{/ruxitagentjs\_}.

However, when Dynatrace wants to deploy its monitoring code on its own site
\url{www.dynatrace.com} (and presumably make sure that it is not blocked by
filter lists) it inlines the entire 203k lines \JS{} library into the header of the
page, preventing existing filter lists from blocking its loading.

\subsection{Combining Code}
Site authors also evade filter lists by combining benign and malicious code into
a single code unit.  This can be done by trivially concatenating \JS{} files
together, or by using popular build tools that combine, optimize
and/or obfuscate many \JS{} files into a single file.

Combining tracking and user-serving code into a single \JS{} unit is difficult
for existing tools to defend against.  Unlike the previous two cases, these
scripts may be easy for filter list maintainers to discover.  However, they
present existing privacy tools with a no-win decision: blocking the script
may prevent the privacy-or-security harm, but break the page for the user;
not blocking the script allows the user to achieve their goals on the site,
though at possible harm to the Web user.

\subsubsection{Classification Methodology}
We identified cases of evasion by ``combing code'' by looking for cases where
the AST of a blocked script is a subgraph of an evaded script,
where both scripts generated the same signature. To detect such
cases, we again use Esprima to generate AST for all scripts that match
the same signatures.  We then look for cases where the AST of a blocked script is fully contained in the AST of an evaded script.
More specifically, if an evaded script's AST has a subtree that is both %
\begin{enumerate*}[label=(\roman*)]
  \item structurally identical to the AST of a blocked script (i.e.,
subtree isomorphism)
  \item the corresponding AST nodes in both trees have the same node type, and
  \item both scripts generated the same signature
\end{enumerate*}%
, we then labeled it as a case of evasion by ``code combining''. In total we observed
\NumEvasionByBundling{} unique scripts (\NumInstanceEvasionByBundling{} instances) that were privacy-and-security harming
scripts combined with other scripts.

\subsubsection{Case Study: Insights JavaScript SDK}
Insights is a \JS{} tracking, analytics and telemetry package from
Microsoft,\footnote{\url{https://docs.microsoft.com/en-us/azure/azure-monitor/overview}}
that allows application developers to track and record visitors.  It includes
functionality identified by \EP{} as privacy-harming, and is blocked by
the filter rule \texttt{||msecnd.net/scripts/a/ai.0.js}.

In order to evade \EP{}, some sites copy the Microsoft Insights code from
the Microsoft provided URL, and included it, among many other libraries, in
a single \JS{} file.  This process is sometimes called ``bundling'' or
``packing''.  As one example, the website \url{https://lindex.com} includes
the Insights library, along with the popular Vue.js and Mustache libraries,
in a single
URL,\footnote{\url{https://lindex.com/web-assets/js/vendors.8035c13832ab6bb90a46.js}}
packaged together using the popular
WebPack\footnote{\url{https://webpack.js.org/}} library.

\subsection{Included Library}
Finally, filter lists are unable to protect against \JS{} code including
common privacy-harming libraries.  Such libraries are rarely, if ever,
included by the site directly, but are instead downstream dependencies by
the libraries directly included on the website.
These cases are common, as \JS{} build systems emphasize small, reusable libraries.
Downstream libraries are difficult for filter lists to target because there is
no URL filter list maintainers can block; instead, filter list maintainers
can only target the diverse and many bespoke \JS{} applications that
include the libraries.

\subsubsection{Classification Methodology}
We identified \NumEvasionByCommonCode{} unique scripts (\NumInstanceEvasionByCommonCode{} instances) in the Alexa 100K that include
such privacy-and-security threatening code as a dependency. These were found by looking for
common significant subtrees between ASTs.  More specifically, when two scripts
generated the same signature, and the AST of the blocked script and the AST of a
not-blocked script, contained significant identical subtrees.  We point out
the possibility for false-positive here, since two scripts generating the
signature might have common AST subtrees that are unrelated to the
privacy-or-security-affecting behavior being signatured.  (e.g., both scripts
could include the jQuery library, but not have that library be the part of
either code unit involved in the signature).

It is difficult to programmatically
quantify the frequency of such false positives due to the complexity of
the \JS{} code involved, which is often obfuscated to deter manual
analysis. Nevertheless, we point out that for scripts in this category, %
\begin{enumerate*}[label=(\roman*)]
  \item our signatures offer considerable improvements over the current
  state-of-the-art, by allowing automatic flagging of scripts that exhibit the
  same privacy-harming semantics as existing blocked scripts, and
  \item we believe these false positives
  to be rare, based on a best-effort, expert human evaluation (we encountered
  only one such case in a human evaluation of over 100 randomly sampled examples,
  performed during the signature size sampling described in Section \ref{sec:meth}).
\end{enumerate*}

\subsubsection{Case Study: Adobe Visitor API}
The ``Visitor API'' is a library built by Adobe, that enables the fingerprinting
and re-identification of site visitors.  It is never included directly
by sites, but is instead included by many other tools, many of which
also generated and sold by Adobe (e.g. Adobe Target).  Some of these
Adobe-generated, Visitor API-including libraries, are blocked by
the \EP{} rule \texttt{||adobetm.com$\hat{~}$\$third-party}.

Other libraries that include the Adobe Visitor API code though are missed
by filter lists, and thus are undefended against.  For example, the
site \url{ferguson.com} indirectly loads the Visitor API code on its site,
through the site's ``setup''
code.\footnote{\url{https://nexus.ensighten.com/ferguson/fergprod/Bootstrap.js}}
There many other similar examples of downstream, privacy-and-security
harming libraries included by diverse \JS{} applications, following this
same pattern.
\section{Discussion}
\label{sec:discussion}

\subsection{Comparison to Hash-Based Detection}

Given the complexity of the signature-based approach presented by this work,
we compared the usefulness of signature-based matching with a much simpler
approach of detecting evasion by comparing code text. More specifically,
we measured how many cases of evasion that we detected by using signatures \emph{would have been missed} by only comparing the text (here, hash) of code units.
We find that the majority of the evasion cases we identify using per-event-loop
signatures would be missed by simple text-comparison approaches.

First, we note the majority of evasions discussed in
Section~\ref{sec:taxonomy} cannot be detected by trivial text-comparison
approaches. For example, a simple technique based on comparing hashes of the
script text against known-bad scripts can only find cases where the exact same
script has been moved \emph{verbatim} from one URL to another, or copied
\textit{verbatim} into a larger code unit; it would fail to find evasion
resulting from even trivial modifications, minification, or processing
by bundling (\EG{} WebPack-like) tools.

Second, we find that our signature-based approach is able to identify a
significant number of cases that text-only approaches would miss. Only 411 of
the \NumUniqeScriptsEvasionByMovedScript{} unique scripts we observed in the
``moving code'' category of our taxonomy
(Section~\ref{sec:taxonomy:methodology}) had identical script text (\IE{}
SHA-256 hash); in the remaining 309 cases the scripts behavior was identical but
the script text was modified (a false negative rate of 42.8\% \emph{in the ``moving code'' category alone}).  However, the
simpler, hash-based approach identified 7,515 of the
\NumInstanceEvasionByMovedScript{} incidents (\IE{} not unique) of moved scripts.
Put differently, a text-comparison approach would correctly handle most
\textit{cases} of scripts being moved, but would miss 42.8\% unique moved
scripts (note that by its nature, a script that has been moved \emph{verbatim}
to another URL is a special case of ``moving code'' in our taxonomy).

Furthermore, evaded scripts in the bundling and common code categories
\emph{cannot} be straightforwardly detected by comparing hashes of the script
text, since by definition these scripts contain new code and thus the hash of
the script will be different. Indeed, it is challenging, if not impossible, to
use text-based detection methods against these evasion techniques. By
comparison, since our approach targets the behavior rather than the delivery
mechanism of the harmful scripts (and regardless of whether they are obfuscated
and/or bundled with other scripts), it can detect evaded scripts whenever their
privacy-harming functionalities are executed.

\subsection{Countermeasures}
\label{sec:discussion:defense}
This work primarily focuses on the problem of measuring how often
privacy-and-security affecting code evades filter lists, by building
behavioral signatures of known-undesirable code, and looking for instances
where unblocked code performs the same privacy-and-security harming behaviors.
In this section we discuss how this same methodology can be used to protect
web users from these privacy-and-security threatening code units.

We consider three exhaustive cases, and possible defenses against each: blocked
code being moved to a new URL, privacy-and-security affecting event-loop turns
that affect storage but not network, and privacy-and-security affecting
event-loop turns that touch network.

\subsubsection{Moved Code}
In cases where attackers evade filter lists by moving code to a new URL,
our approach can be used to programmatically re-identify those moved code units,
and generate new filter lists rules for the new URLs. Using this approach,
we have generated \NumFilterListRules{} new filter list URLs, compatible with
existing popular content blocked tools like AdBlock Plus and uBlock Origin.
Further, we have submitted many of these new filter list rules to the
maintainers \EL{} and \EP{}; many have been upstreamed, and many more
are being reviewed by the maintainers.

\subsubsection{Event-Loop Turns Without Network}
Instances of code being inlined, or privacy-or-security affecting code being
combined with other code, are more difficult to defend against, and require
runtime modifications.  These have \textit{not} been implemented as part of
this work, but we discuss possible approaches for doing so here.\footnote{These
are not abstract suggestions either; we are working with a privacy-focused
browser vendor to implement these proposals.}  We note that the single-threaded
model of the browser means that signature-matching state only needs to be
maintained per \JS{} content, to track the behavior of the currently executing
script; state does not need to be maintained per code unit.

In cases where the privacy-harming, event-loop signature only consists of
storage events (i.e. no network behavior), we propose staging storage for the
length of each event-loop turn, discarding the storage modifications if the
event-loop turn matches a signature, and otherwise committing it. This would
in practice be similar to how Safari's ``intelligent tracking protection''
system stages storage until the browser can determine if the user is being
bounce-tracked.

\subsubsection{Event-Loop Turns With Network}
The most difficult situation for our signature-based system to defend against
is when the event-loop turn being ``signatured'' involves network activity,
as this may force a decision before the entire signature can be matched (i.e.
if the network event occurs in the middle of signature).  In such cases,
runtime matching would need to operate, on average, with half as much
information, and thus would not provide protection in 50\% of cases.  While
this is not ideal, we note that this is a large improvement over current privacy
tools, which provide no protections in these scenarios.  We leave
ways of more comprehensively providing runtime protects against
network-involving, security-and-privacy harming event-loop turns as future
work.

\subsection{Accuracy Improvements}
This work generates signatures from known privacy-and-security harming scripts,
using as input the behaviors discussed in Table
\ref{table:signature-events}.  While we have found these signatures
to be highly accurate (based on the methodology discussed in Section
\ref{sec:results} and the AST-based matching discussed in Section
\ref{sec:taxonomy}), there are ways the signatures could be further improved.
First, the signatures could be augmented with further instrumentation points, to
further reduce any false positives, and build even more unique signatures
per event-loop turn. Second, we expect that for many scripts, calling a
given function will result in neither purely deterministic behavior, nor
completely unpredictable behavior; that some subsections of code can
result in more than one, but less than infinite, distinct signatures. Further
crawling, therefore, could increase recall by more comprehensively generating
all possible signatures for known privacy-and-security affecting code.

\subsection{Web Compatibility}
Current web-blocking tools suffer significant trade-offs between coverage
and usability. Making decisions at the URL level, for example, will result
in cases of over blocking (and breaking the benign parts of a website)
or under blocking (and allowing the privacy harming behavior). By moving
the unit of analysis to the event-loop turn, privacy tools could make finer
grained decisions, and do a better job distinguishing between unwanted
and benign code.  While we leave an evaluation of the web-compatibility
improvements of our proposed per-event-loop-turn system to further work,
we note it here as a promising direction for researchers and activists looking
to make practical, usable web privacy tools.

\subsection{Limitations}
\label{sec:discussion:limitiations}
Finally, we note here limitations of this work, and suggestions for how
they could be addressed by future work.

\subsubsection{Automated Crawling}
The most significant limitation
is our reliance on automated-crawls to build signatures. While such automated
crawls are useful for covering a large portion of the web, they have significant
blind spots, including missing scripts only accessible after authentication
on a site, or only after performing complex interactions on a page.  Prior
work has attempted to deal with this though paid research-subject
volunteers~\cite{snyder2017most}, or other ways of approximating real world
usage. Such efforts are beyond the scope of this project, but we note
them here for completeness.

\subsubsection{Evasion}
Second, while our behavioral-based signature approach is far more robust to
evasion than existing URL focused web privacy-and-security tools, there are
still cases where the current approach could be fooled. For example, if
an attacker took a privacy-harming behavior currently carried out by a single
script, and spread the functionality across multiple colluding code units,
our system would not detect it (though it could with some post-processing
of the graph to merge the behavior of colluding scripts).  Similarly, attackers
might introduce intentional non-determinism in their code, by, for example,
shuffling the order of some operations in a way that does not affect the
code's outcome.

While our system could account for some of these cases through
further crawling (to account for more possible code paths) or generalizations
in signature generation, we note this attack as a current limitation and area
for future work.

We note, however, that our signature-based approach would be robust to many
forms of obfuscation that would confuse other signature-based approaches.
Because our approach relies on code's behavior, and not text representation,
our approach is resilient against common obfuscation techniques like code
rewriting, modifying the text's encoding, and encrypting the code.  We also
note that our approach would not be fooled by obfuscation techniques that
only changed control flow \textit{without also changing script behavior};
our technique would be robust against obfuscation techniques that only
modify \JS{} structure.

\subsubsection{False Positives}
Our approach, like all signature-based approaches, makes trade offs between
false-positive and false-negative rates. Encoding more information in a
signature increases the confidence in cases where the signature matches
observed behavior, but at the risk of missing more similar-but-not-identical
cases. As described in Section~\ref{sec:meth:privacy-sigs}, our system
only builds signatures for graphs including at least \SignatureMinEdges{} edges
and at least \SignatureMinNodes{} nodes. This minimum graph size was selected
by iteratively increasing the minimum graph size until we no longer observed
any false positives through manual examination.

However, it is possible that despite the above described process, our minimum
signature size is not sufficient to prevent some false positives; given the
number and diversity of scripts on the web, it is nearly a certainty that there
are instances of both benign and undesirable code that perform the same
\SignatureMinEdges{} behaviors, interacting with \SignatureMinNodes{} similar
page structures, even if we observed no such instances in our
manual evaluation. Deployments of our work that prefer accuracy
over recall could achieve such by increasing the minimum graph size
used in signature generation.
\section{Related Work}
\label{sec:relwork}

\subsection{Blocking trackers}
The current line of defense that most users have against web tracking is via
browser extensions~\cite{adblockplus, ublock, ghostery, disconnect}. These
extensions work by leveraging hand-crafted filter lists of HTML elements and
URLs that are connected with advertisers and trackers~\cite{easylist}. There are
also dynamic approaches for blocking, like Privacy Badger from
EFF~\cite{privadger}, which tracks images, scripts and advertising from third
parties in the visited pages and blocks them if it detects any tracking
techniques. The future of browser extensions as web tracking prevention tools is
currently threaten by the transition to the newest version of WebExtensions
Manifest v3~\cite{webextensionsv3}, which limits the capabilities of dynamically
making decisions to block content.

Previous research has also focused on automated approaches to improve content
blocking. Gugelmann~\etal, built a classifier for identifying privacy-intrusive
Web services in HTTP traffic~\cite{gugelmann2015automated}. NoMoAds leverages
the network interface of a mobile device to extract features and uses a
classifier to detect ad requests~\cite{shuba2018nomoads}.


\subsection{Instrumenting the browser}
Extracting information from the browser is mandatory step into understanding web
tracking. Previous approaches, like OpenWPM, have focused on leveraging a
browser extension to monitor the events of a visited
page~\cite{englehardt2016census}. In-band approaches like OpenWPM inject JS into
the visited page in order to capture all events, which can affect their
accuracy, as they are running at the same level as the monitored code. Recently,
we have observed a shift in pushing more browser instrumentation out-of-band
(in-browser)~\cite{iqbal2020adgraph,li2018jsgraph,vv8-imc19}. In this paper, we
follow a similar out-of-band approach, where we build signatures of tracking
scripts based on the dynamic code execution by instrumenting Blink and V8 in the
Chromium browser.

\subsection{Code Similarity}
Code similarity is a well-established research field in the software engineering
community~\cite{roy2007survey}. From a security perspective, finding code
similarities with malicious samples has been explored in the past.
Revolver~\cite{revolver-UsenixSec13} performed large-scale clustering of
JavaScript samples in order to find similarities in cases where the
classification is different, automatically detecting this way evasive samples.

Ikram~\etal~\cite{ikram2017towards}, suggested the use of features from JavaScript programs using
syntactic and structural models to build a classifier that detects scripts with
tracking~\cite{ikram2017towards}. Instead of relying on syntactic and structural
similarity, in our work we identify tracking scripts based on the tracking
properties of their execution in the browser, defeating this way techniques like
obfuscation~\cite{skolka2019} and manipulation of
ASTs~\cite{fass2019hidenoseek}.

\subsection{Other Content Blocking Strategies}

Another approach to block content is via perceptual detection of
advertisements~\cite{storey2017future,sentinel}. This approach is based on the
identifying advertisements based on known visual patterns that they have, such
as the AdChoices standard~\cite{adchoices}. Although this is an exciting new
avenue of blocking content on the web, there is already work that aims to create
adversarial attacks against perceptual ad blocking~\cite{adverceptual}.

\section{Conclusion}
\label{sec:conclusion}

The usefulness of content blocking tools to protect Web security and privacy 
is well understood and popularly enjoyed. However, the URL-focus of these
tools means that the most popular and common tools have trivial circumventions,
which are also commonly understood, though frequently ignored for lack of
alternatives.

In this work we make several contributions to begin solving this problem,
by identifying malicious code using highly granular, event-loop turn level
signatures of runtime \JS{} behavior, using a novel system of browser
instrumentation and graph-based signature generation. We contribute not
only the first Web-scale measurement of how much evasion is occurring on the
Web, but also the ground work for practical defenses. To further contribute
to the goal a privacy-and-security respecting Web, we also contribute 
the source code for our instrumentation and signature generation systems,
the raw data gathered during this work, and filter list rules that can
help users of existing tools defend against a subset of the
problem~\cite{semanticsignatures}.

\bibliographystyle{plain}
\bibliography{main}


\end{document}